\documentclass[prd,twocolumn,preprintnumbers,amsmath,amssymb,superscriptaddress,floatfix,nofootinbib]{revtex4-2}
\usepackage{graphicx} 
\usepackage{dcolumn}  
\usepackage{bm}       
\usepackage{float}
\usepackage{hyperref}
\hypersetup{
    colorlinks=true,
    linkcolor=blue,
    anchorcolor=blue,
    citecolor=blue,
    urlcolor=blue 
}
\newcommand{\sm}[1]{\scriptscriptstyle{#1}}

\begin{document}

\title{Mass spectrum and decay widths of charmonium-like mesons: A diabatic approach with complex scaling
}
\author{Zi-Zhao Zhang}\author{Bo-Chao Liu}
\email{liubc@xjtu.edu.cn}
\affiliation{MOE Key Laboratory for Nonequilibrium Synthesis and Modulation of Condensed Matter, School
of Physics, Xi’an Jiaotong University, Xi’an 710049, China.}

\begin{abstract}
In this work, we extend our previous diabatic framework for the charmonium-like spectrum below $4.3$ GeV by introducing the complex scaling method. Unlike the previous framework, the present approach explicitly incorporates couplings to the meson-meson continuum, comprehensively accounting for its contributions to the physical states. Consequently, it allows bound and resonant states to be treated on an equal footing, enabling the direct extraction of decay widths from complex energy eigenvalues without introducing any additional free parameters. Using the obtained solutions, we calculate the channel weights and complex root-mean-square radii ($r_{\sm{RMS}}$) to elucidate their internal structures. Specifically, we find that the $\chi_{c1}(3872)$, $\psi(4040)$, and $\psi(4230)$ states exhibit significant molecular characteristics. Furthermore, we propose the $X(3940)$ as a candidate for a $J^{PC}=1^{++}$ state and discuss the nature of the $\chi_{c0}(2P)$ resonance in detail. 
\end{abstract}

\maketitle
\section{INTRODUCTION}

In recent years, the experimental observation of numerous charmonium-like states has posed serious challenges to the conventional $q\bar{q}$ description of charmonium \cite{ParticleDataGroup:2024cfk}. A prominent example is the $X(3872)$ \cite{Belle:2003nnu}, which has a mass extremely close to the $D^0\bar{D}^{*0}$ threshold and exhibits properties that are difficult to reconcile with a pure $c\bar{c}$ assignment. Following its discovery, an increasing number of XYZ states have been reported, such as the $\chi_{c0}(3860)$ \cite{Belle:2017egg,Yu:2017bsj}, $X(3940)$ \cite{Belle:2005lik,LHCb:2024vfz}, $\chi_{c2}(3930)$ \cite{Belle:2005rte,LHCb:2020pxc}, and others \cite{Belle:2004lle,BESIII:2023tll,BES:1999wbx,LHCb:2021uow}. Many of these states exhibit masses, decay widths, or decay patterns that deviate markedly from the conventional quark model predictions. These experimental findings indicate that the simple Cornell-type potential, while highly successful for low-lying charmonium states \cite{Eichten:1974af,PhysRevD.17.3090,Godfrey:1985xj,barnes2005higher,eichten2008quarkonia}, is insufficient to provide a unified description of highly excited or near-threshold states \cite{Belle:2003nnu,Belle:2005lik,Belle:2005rte,LHCb:2020bls, LHCb:2021uow,LHCb:2022aki,Brambilla:2019esw,Chen:2022asf}. Consequently, additional dynamical mechanisms beyond the conventional $q\bar{q}$ picture must be taken into account.

A notable feature of many newly observed charmonium-like states is their proximity to open-charm meson-meson thresholds. This proximity implies that coupled-channel effects, induced by virtual meson loops, can play a crucial role. Therefore, a reliable description of these states requires the inclusion of meson-meson channels. To this end, various theoretical approaches incorporating coupled-channel dynamics have been developed (see, e.g., Refs.~\cite{PhysRevD.80.014012,PhysRevD.72.034010,PhysRevD.76.077502,ono1984continuum}). Among these, an attractive strategy is the adiabatic approach.

The adiabatic approximation, often referred to as the Born-Oppenheimer (B-O) approximation, was originally developed in molecular physics and has been widely applied to atomic and molecular physics, as well as to heavy-quark systems within QCD \cite{born1985quantentheorie,bali2001qcd,castella2022heavy,Brambilla:2026xxx}. In its standard formulation, the B-O approach assumes both adiabaticity and a single-channel configuration. However, for charmonium-like states located near open-flavor meson-meson thresholds, the mixing between $Q\bar{Q}$ and meson-meson configurations becomes significant, rendering the single-channel assumption inadequate \cite{PhysRevD.102.074002}.
In such scenarios, a proper description requires a set of coupled-channel Schr\"odinger equations supplemented by nonadiabatic coupling terms, whose explicit treatment is technically challenging~\cite{beyondBorn-Oppenheimer}. A natural way to overcome this difficulty is to generalize the B-O framework and adopt the diabatic formalism, wherein the configuration mixing is encoded in a matrix-valued potential within a multichannel Schr\"odinger equation.

In this context, recent unquenched lattice QCD calculations play a pivotal role by providing first-principles results for the static energies of heavy quark-antiquark sources in the presence of open-flavor meson-meson channels \cite{Bali:2005fu,Bulava_2019}. These lattice results can be directly mapped to the elements of the diabatic potential matrix, thereby establishing a concrete connection between fundamental QCD and effective multichannel descriptions of quarkonium-like states. This lattice-informed diabatic potential approach has already been successfully applied to bottomonium-like systems and has recently been extended to incorporate systematic treatments of multiple meson-meson channels \cite{Bruschini:2021yul,Bruschini:2021uql,Bruschini:2023wpo,Bruschini:2024fyj,Braaten:2024tbm,Berwein:2024ztx,Bruschini:2023wpo}.

The application of the diabatic approach to calculating charmonium-like mass spectra has been extensively explored in previous studies \cite{PhysRevD.102.074002,Lebed:2022vks,Lebed:2024rsi,Bruschini_2021,Zhang:2025bex}. Since this method essentially relies on solving a coupled-channel Schr\"odinger equation for bound states, additional approximations are typically required to address continuum effects and decay properties. A common strategy is to first neglect the open-channel continuum contributions to obtain a discrete bound-state spectrum. The effects of the continuum channels are subsequently incorporated perturbatively to shift the masses of the resulting states and estimate their corresponding decay widths \cite{Bruschini_2021}.

The complex scaling method (CSM) provides a unified framework for describing bound states, resonances, and continuum states\cite{Aguilar:1971ve,Balslev:1971vb}. 
Under the complex scaling transformation, the coordinates and their conjugate momenta are rotated as $r\to r e^{i\theta}$ and $p\to p e^{-i\theta}$, respectively, and the Hamiltonian is transformed into a complex-scaled Hamiltonian. In this representation, resonant states appear as discrete eigenstates with complex eigenvalues, separated from the rotated continuum spectrum in the complex-energy plane. For an eigenvalue written as $E=M-i\Gamma/2$, the real part gives the resonance mass, while the decay width is obtained as $\Gamma=-2\,\mathrm{Im}\,E$. The CSM has been widely applied to the study of hadron resonances \cite{Wu:2024ocq,Oka:2019mrd,Song:2024ngu}.

Given that the CSM is well suited for treating coupled-channel problems, it is natural to ask whether it can be incorporated into the diabatic framework, where the dynamics are governed by a multichannel Schrödinger equation with matrix-valued potentials. In our previous work \cite{Zhang:2025bex}, we investigated the charmonium-like spectrum within the diabatic framework, explicitly including spin-dependent interactions. However, the contributions from meson–meson continuum channels were neglected, so decay processes and their corresponding widths were not evaluated. In the present work, we extend this approach by integrating the CSM into the diabatic framework. Without introducing additional free parameters, this unified treatment provides a more accurate description of the spectrum, naturally characterizes both bound and resonant states (the latter emerging as isolated complex eigenvalues of the non-Hermitian Hamiltonian), and enables the direct calculation of resonance decay widths.

The paper is organized as follows. In Sec.~II, we introduce the theoretical framework, including a brief overview of the diabatic approach, the construction of the potential matrix, and an outline of the CSM. In Sec.~III, we present our numerical results and discuss their physical implications. Finally, we summarize our findings in Sec.~IV.

\section{Framework}
\subsection{Diabatic Formalism}
To describe the heavy quarkonium system, we adopt the Hamiltonian formalism as detailed in Ref.~\cite{PhysRevD.102.074002}. In the center-of-mass frame of $Q\bar{Q}$, the Hamiltonian is expressed as:
\begin{equation}
H = \frac{\boldsymbol{p}^2}{2\mu} + H^{\mathrm{light}},
\label{eq:hamiltonian}
\end{equation}
where $\mu$ denotes the reduced mass of the $Q\bar{Q}$ system, $\boldsymbol{p}$ is the $Q\bar{Q}$ relative momentum, and $H^{\mathrm{light}}$ governs the dynamics of the light degrees of freedom (light quarks and gluons), including their interactions with $Q\bar{Q}$.

Since the mass of the heavy quark is significantly larger than $\Lambda_{\mathrm{QCD}}$ (the characteristic energy scale of the light degrees of freedom), the adiabatic approximation is well justified. In this limit, the motion of the heavy quark pair is neglected when determining the light-field dynamics. Consequently, the relative separation of the heavy quarks, $\boldsymbol{r}$, is treated as a fixed parameter.

The dynamics of the light fields are then described by the following eigenvalue equation for a fixed $\boldsymbol{r}$:
\begin{equation}
H^{\mathrm{light}}_{\mathrm{static}}(\boldsymbol{r}) |\zeta_i(\boldsymbol{r})\rangle = V_i(\boldsymbol{r}) |\zeta_i(\boldsymbol{r})\rangle,
\label{eq:adiabatic_eq}
\end{equation}
where $|\zeta_i(\boldsymbol{r})\rangle$ denotes the $i$-th eigenstate of the light fields at separation $\boldsymbol{r}$. The corresponding eigenvalue, $V_i(\boldsymbol{r})$, represents the energy of the light fields at the fixed $Q\bar{Q}$ separation, which can be calculated \textit{ab initio} in lattice QCD.

\begin{figure}[htbp]
    \centering
    \includegraphics[width=\columnwidth]{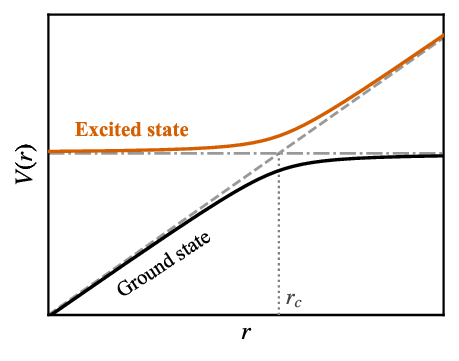}
    \caption{Schematic representation of the string breaking mechanism. The dashed and dash-dotted curves depict the ground-state static energy and the meson-meson threshold, respectively. The solid lines represent the ground and excited-state static energies, which exhibit an avoided crossing. The crossing point $r_c$ is marked.}
    \label{fig:string_breaking}
\end{figure}

Unquenched lattice QCD calculations~\cite{Bali:2005fu,Bulava_2019} have demonstrated that, due to string breaking, an \textit{avoided crossing} occurs in the static energy spectrum at intermediate heavy quark separations. This phenomenon indicates a strong mixing between the bare $Q\bar{Q}$ configuration and the meson-meson configuration. A schematic illustration of this behavior for a $Q\bar{Q}$ system coupled to a single meson-meson threshold is presented in Fig.~\ref{fig:string_breaking}.

To solve the Schr\"odinger equation $(H-E)|\Psi\rangle = 0$, we employ the diabatic expansion. The total wave function $|\Psi\rangle$ is expanded in terms of the light-field eigenstates $|\zeta_i(\mathbf{r}_0)\rangle$ defined at a fixed $Q\bar{Q}$ separation $\mathbf{r}_0$:
\begin{equation}
|\Psi\rangle = \sum_i \int d\mathbf{r}' \, \psi_i(\mathbf{r}', \mathbf{r}_0) \, |\mathbf{r}'\rangle |\zeta_i(\mathbf{r}_0)\rangle,
\label{eq:diabatic_expansion}
\end{equation}
where $|\mathbf{r}'\rangle$ is the position eigenstate of the heavy quarks, and $\psi_i(\mathbf{r}', \mathbf{r}_0)$ is the corresponding expansion coefficient. Here, $\psi_i(\mathbf{r}', \mathbf{r}_0)$ represents the wave function governing the relative motion of the heavy quarks for the $i$-th light-field configuration.

The choice of  $\mathbf{r}_0$ is critical for the physical interpretation of the basis states. As illustrated in Fig.~\ref{fig:string_breaking}, the mixing between the $Q\bar{Q}$ and meson-meson configurations is strongly localized near the avoided crossing point $r_c$. When $\mathbf{r}_0$ is chosen sufficiently far from $\mathbf{r}_c$, the eigenstates $|\zeta_i(\mathbf{r}_0)\rangle$ correspond to pure configurations, namely the bare $Q\bar{Q}$ or meson-meson states. Under this condition, the ground state $|\zeta_0(\mathbf{r}_0)\rangle$ corresponds to the pure quarkonium configuration, while the $i$-th excited state $|\zeta_i(\mathbf{r}_0)\rangle$ ($i \ge 1$) corresponds to the $i$-th meson-meson channel ($M\bar{M}_i$).

By substituting Eq.~(\ref{eq:diabatic_expansion}) into the Schrödinger equation and projecting onto the basis states $\langle \mathbf{r} | \langle \zeta_j(\mathbf{r}_0) |$, we arrive at the following coupled-channel equations:
\begin{equation}
\sum_i \left( -\frac{\hbar^2}{2\mu}\delta_{ji}\nabla^2 + V_{ji}(\mathbf{r}, \mathbf{r}_0) - E\delta_{ji} \right) \tilde{\psi}_i(\mathbf{r}, \mathbf{r}_0) = 0.
\label{eq:coupled_eq}
\end{equation}
The interaction matrix elements $V_{ji}$, which play a central role in our calculation, are defined as the expectation values of the light-field Hamiltonian:
\begin{equation}
V_{ji}(\mathbf{r}, \mathbf{r}_0) \equiv \langle \zeta_j(\mathbf{r}_0) | H^{\mathrm{light}}(\mathbf{r}) | \zeta_i(\mathbf{r}_0) \rangle.
\label{eq:potential_matrix}
\end{equation}

The diagonal element for the heavy quark core takes the form of the standard Cornell potential:
\begin{equation}
V_{c\bar{c}}(r) \equiv \langle \zeta_{Q\bar{Q}} | H^{\mathrm{light}}(\mathbf{r}) | \zeta_{Q\bar{Q}} \rangle.
\label{eq:V_conf}
\end{equation}

The interactions within the meson-meson channels are described by:
\begin{equation}
V^{M\bar{M}}_{ji}(r) \equiv \langle \zeta_{M\bar{M}_j} | H^{\mathrm{light}}(\mathbf{r}) | \zeta_{M\bar{M}_i} \rangle,
\label{eq:V_MM}
\end{equation}
where $1 \le i,j \le N$. 

Finally, the off-diagonal terms responsible for string breaking are identified as the mixing potentials:
\begin{equation}
V^{\mathrm{mix}}_j(r) \equiv \langle \zeta_{Q\bar{Q}} | H^{\mathrm{light}}(\mathbf{r}) | \zeta_{M\bar{M}_j} \rangle,
\label{eq:V_mix}
\end{equation}
where we identify $V^{\mathrm{mix}}_j(r)$ with $V_{0j}(\mathbf{r}, \mathbf{r}_0)$. 

With the notation established above, the coupled-channel equations can be recast into a compact matrix form:
\begin{equation}
\label{eq:matrix_SE}
(\mathbf{K} + \mathbf{V}(r)) \mathbf{\Psi}(r) = E \mathbf{\Psi}(r).
\end{equation}
Here, $\mathbf{K}$, $\mathbf{V}(r)$, and $\mathbf{\Psi}(r)$ represent the kinetic energy matrix, the potential energy matrix, and the multi-component wave function vector, respectively.

The kinetic energy matrix $\mathbf{K}$ is diagonal, where the diagonal elements represent the kinetic energy operators of each independent channel:
\begin{equation}
\mathbf{K} = \mathrm{diag}(-\frac{\hbar^2}{2\mu_{c\bar{c}}}\nabla^2, -\frac{\hbar^2}{2\mu^{(1)}_{M\bar M}}\nabla^2, \dots, -\frac{\hbar^2}{2\mu^{(N)}_{M\bar M}}\nabla^2)
\end{equation}
where $\mu_{c\bar c}$ and $\mu_{M\bar{M}}^{(n)}$ denote the reduced masses of the $c\bar{c}$ core and the $n$-th meson-meson pair ($1 \le n \le N$).

For the interaction matrix $\mathbf{V}(r)$, we adopt the approximation used in lattice QCD studies~\cite{Bulava_2019} by neglecting the direct couplings between different meson-meson channels (i.e., $V_{ij} = 0$ for $i \ne j$ with $i,j \ge 1$). Consequently, the potential matrix is constructed as:
\begin{equation}
\mathbf{V}(r) =
\begin{bmatrix}
V_{c\bar{c}}(r) & V^{\mathrm{mix}}_1(r) & \cdots & V^{\mathrm{mix}}_N(r) \\
V^{\mathrm{mix}}_1(r) & V^{M\bar{M}}_1(r) & &  \\
\vdots & & \ddots & \\
V^{\mathrm{mix}}_N(r) &  & & V^{M\bar{M}}_N(r)
\end{bmatrix}.
\end{equation}

The total wave function $\mathbf{\Psi}(\mathbf{r})$ comprises the channel wave functions for all considered channels:
\begin{equation}
\mathbf{\Psi}(\mathbf{r})=\left( \psi_{c\bar c}(\mathbf{r}), \psi_{M\bar M}^{(1)}(\mathbf{r}), \dots, \psi_{M\bar M}^{(N)}(\mathbf{r}) \right)^T,
\end{equation}
which satisfies the normalization condition
\begin{equation}
\int d^3\mathbf{r} \mathbf{\Psi}^\dagger(\mathbf{r}) \mathbf{\Psi}(\mathbf{r}) = 1.
\end{equation}

\subsection{Potential Matrix with Spin-dependent Terms}

The interaction between the charm quark and antiquark, $V_{c\bar{c}}(r)$, is modeled by a nonrelativistic potential including both a spin-independent central part and spin-dependent corrections \cite{Lucha:1995zv,PhysRevD.29.110}. Explicitly, the total potential is written as a sum of four terms:
\begin{equation}
V_{c\bar{c}}(r) = V_{\mathrm{conf}}(r) + V_{SS}(r) + V_{SO}(r) + V_{T}(r).
\label{eq:total_cc_potential}
\end{equation}

The first term, $V_{\mathrm{conf}}(r)$, represents the standard Cornell potential~\cite{eichten2008quarkonia,bali2001qcd}, which dominates the spin-averaged spectrum. It consists of a short-range color Coulomb interaction and a long-range linear confining potential:
\begin{equation}
V_{\mathrm{conf}}(r) = -\frac{4}{3}\frac{\alpha_s}{r} + b r,
\label{eq:cornell}
\end{equation}
where $\alpha_s$ is the strong coupling constant and $b$ is the string tension parameter.

The remaining terms in Eq.~(\ref{eq:total_cc_potential}) account for the spin-dependent splittings, which are derived from the one-gluon exchange (OGE) interaction in the Breit-Fermi limit. To regularize the singularity of the contact interaction, we employ a Gaussian-smeared form for the spin-spin (hyperfine) interaction~\cite{barnes2005higher}:
\begin{equation}
V_{SS}(r) = \frac{32\pi\alpha_s}{9m_c^2} \left( \frac{\sigma}{\sqrt{\pi}} \right)^3 e^{-\sigma^2 r^2} \mathbf{S}_c \cdot \mathbf{S}_{\bar{c}},
\label{eq:V_SS}
\end{equation}
where $m_c$ is the charm quark mass, $\sigma$ characterizes the smearing width, and $\mathbf{S}_c$ and $\mathbf{S}_{\bar{c}}$ denote the spin operators of the charm and anti-charm quarks, respectively.

The fine structure of the spectrum is governed by the spin-orbit ($V_{SO}$) and tensor ($V_{T}$) potentials. These terms are taken from Ref.\cite{barnes2005higher}:
\begin{align}
V_{SO}(r) &= \frac{1}{m_c^2} \left( \frac{2\alpha_s}{r^3} - \frac{b}{2r} \right) \mathbf{L} \cdot \mathbf{S}, \label{eq:V_SO} \\
V_{T}(r) &= \frac{4\alpha_s}{m_c^2 r^3} \hat{T}, \label{eq:V_T}
\end{align}
where $\mathbf{L}$ denotes the orbital angular momentum operator and $\mathbf{S} = \mathbf{S}_c + \mathbf{S}_{\bar{c}}$ represents the total spin operator. The tensor operator is defined as $\hat{T} = 3(\mathbf{S}_c \cdot \hat{r})(\mathbf{S}_{\bar{c}} \cdot \hat{r}) - \mathbf{S}_c \cdot \mathbf{S}_{\bar{c}}$.

We calculate the matrix elements of these potentials in the $|J, L, S\rangle$ basis. The expectation values of the relevant spin operators are evaluated as follows:
\begin{subequations}
\begin{align}
\langle \mathbf{S}_c \cdot \mathbf{S}_{\bar{c}} \rangle &= \frac{1}{2}S(S+1) - \frac{3}{4}, \\
\langle \mathbf{L} \cdot \mathbf{S} \rangle &= \frac{1}{2} [J(J+1) - L(L+1) - S(S+1)], \\
\langle \hat{T} \rangle &= 
\begin{cases} 
-\frac{L}{2(2L+3)}, & J=L+1 \\
+\frac{1}{6}, & J=L \\
-\frac{L+1}{2(2L-1)}, & J=L-1 
\end{cases}.
\end{align}
\end{subequations}
Note that the off-diagonal contributions of the tensor term are negligible and thus ignored in this study~\cite{PhysRevD.29.110}. The numerical values of the model parameters $\alpha_s$, $b$, $\sigma$, and $m_c$ are provided in Sec.~\ref{results}A.

For the meson-meson sector, we adopt the free-meson approximation, neglecting interactions between the open-flavor mesons as well as the direct couplings between different meson-meson channels. Under this assumption, the potential matrix elements for the decay channels are diagonal and constant, determined solely by the meson-meson thresholds:
\begin{equation}
V_{M\bar{M}}^{(n)}(r) = T_n \equiv m_{M_1}^{(n)} + m_{\bar{M}_2}^{(n)},
\label{eq:threshold_potential}
\end{equation}
where $m_{M_1}^{(n)}$ and $m_{\bar{M}_2}^{(n)}$ denote the masses of the mesons in the $n$-th channel. We note that while this approximation correctly describes the long-range meson-pair limit, a complete QCD-constrained B–O potential would also include a repulsive color-octet Coulomb term at short distances~\cite{Berwein:2024ztx,Braaten:2024tbm}, which is not considered here.

The coupling between the confined charmonium core and the $n$-th continuum channel is governed by the mixing potential $V_{\mathrm{mix}}^{(n)}(r)$. We adopt the parametrization extracted from lattice QCD studies~\cite{Bulava_2019} and employed in Ref.~\cite{PhysRevD.102.074002}, which models this interaction as a localized function peaking at the string-breaking distance.

Physically, the mixing is strongest at the crossing radius $r_c^{(n)}$, defined as the separation at which the potential energy of the $c\bar{c}$ pair equals the mass of the meson pair, i.e., $V_{c\bar{c}}(r_c^{(n)}) = T_n$. Far from this region, the transition probability is suppressed. Accordingly, the mixing potential is parametrized as a Gaussian function of the energy difference:
\begin{equation}
V_{\mathrm{mix}}^{(n)}(r) = \frac{\Delta}{2} \exp\left\{ -\frac{(V_{\mathrm{c\bar{c}}}(r) - T_n)^2}{2(b\rho)^2} \right\}.
\label{eq:mixing_potential}
\end{equation}
Here, $\Delta$ represents the mixing strength, and $\rho$ characterizes the width of the transition region. The parameter $b$ is the string tension from Eq.~(\ref{eq:cornell}).

It is worth noting that the string breaking phenomenon occurs in the long-range confinement region ($r \sim 1$ fm or larger). Since the spin-dependent potentials derived from one-gluon exchange are short-range interactions, their contribution to the mixing mechanism is negligible. Therefore, in Eq.~(\ref{eq:mixing_potential}), we use only the spin-independent central potential $V_{\mathrm{conf}}(r)$ (the Cornell potential) to determine the radial dependence of the mixing potential. The parameters $\Delta$ and $\rho$ are fixed in the following section.

\subsection{Complex scaling method}
To enable a self-consistent description of resonances in the study of charmonium-like states, we employ the CSM \cite{Aguilar:1971ve,Balslev:1971vb}. In the CSM, the relative coordinate $r$ and the conjugate momentum $p$ are transformed into the complex plane via a transformation $U(\theta)$. This transformation is explicitly defined as:
\begin{equation}
    \label{eq:csm1}  
    U(\theta)rU^{-1}(\theta) = re^{i\theta}, U(\theta)pU^{-1}(\theta) = pe^{-i\theta}
\end{equation}
Under this transformation, the original Hamiltonian $H$ becomes the complex-scaled non-Hermitian Hamiltonian $H_\theta = U(\theta) H U^{-1}(\theta)$, and the corresponding Schr\"odinger equation takes the form:
\begin{equation}
    H_\theta \Psi_\theta = E_\theta \Psi_\theta 
    \label{eq:csm2}
\end{equation}
where $E_\theta$ is complex-valued. In the present work, we adopt the Gaussian Expansion Method (GEM) \cite{Hiyama:2003cu,Hiyama:2018ivm} to solve this equation numerically. 

\begin{figure}[htbp]
    \centering
    \includegraphics[width=\columnwidth]{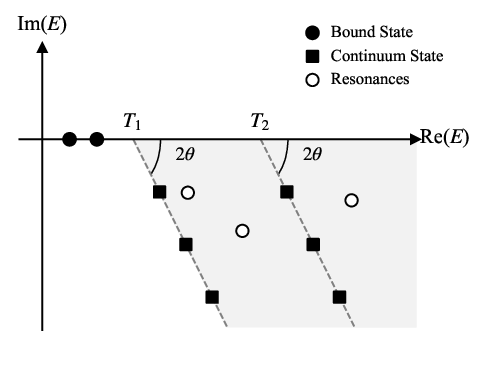}
    \caption{\label{fig:csm_spectrum} Schematic of the eigenvalue distribution of the complex-scaled Hamiltonian $H_\theta$ in the coupled-channel two-body system. $T_i$ represents the $i$-th threshold. }
\end{figure}

In Fig.~\ref{fig:csm_spectrum}, we illustrate the typical eigenvalue distribution in the complex energy plane. According to the ABC theorem \cite{Aoyama:2006hrz}, the continuum states are rotated clockwise by $2\theta$ around the thresholds. In contrast, the discrete solutions—comprising both bound and resonant states—are independent of $\theta$. Specifically, bound states remain on the real axis below the relevant thresholds, whereas resonances appear as isolated poles within the wedge formed by the rotated continua. 
In this framework, these resonance poles correspond to discrete eigenvalues of the complex-scaled Hamiltonian. The physical mass $M$ and decay width $\Gamma$ are then determined from the complex eigenvalue of the identified resonance:
\begin{equation}
    E_\theta = M - i\frac{\Gamma}{2}.
    \label{eq:csm3}
\end{equation}

In addition to the eigenenergies, the CSM provides access to the complex-scaled wave functions. Since the complex-scaled Hamiltonian is non-Hermitian, we employ the c-product (or biorthogonal product) to define the norm and expectation values \cite{Song:2024ngu,Lin:2023ihj,Song:2025xnx,Wu:2024qve,Myo:2021xqp}. Under this definition, the inner product is defined by the square of the wave function rather than the modulus squared. Specifically, the bra vector is not the Hermitian conjugate of the ket-vector. For the coupled-channel system, we calculate the root-mean-square (RMS) radius $r_{\text{RMS}}$ and the component weight $P_i$ for each channel.

The component weight $P_i$ of the $i$-th channel is defined as:
\begin{equation}
\label{eq:c_component_prop}
P_i = (\psi_{i}|\psi_{i}) = \int \psi_{i}(\mathbf{r})^2 d^3\mathbf{r},
\end{equation}
where $\psi_i(\mathbf{r})$ represents the wave function of the $i$-th channel. The total wave function satisfies the normalization condition:
\begin{equation}
\label{eq:c_normalization}
\sum_{i} (\psi_{i}|\psi_{i}) = 1.
\end{equation}
For resonant states, the squared amplitudes are generally complex, and consequently the corresponding component weights are also complex quantities. Although the sum of all component weights is normalized to unity, an individual component \(P_i\) is not necessarily real or positive. When its imaginary part is small, the real part of \(P_i\) may be regarded as an approximate indicator of the relative importance of the corresponding channel, in analogy with the probabilistic interpretation of components in bound states. The imaginary part is associated with the non-Hermitian nature of the resonant-state description. Nevertheless, its detailed physical meaning remains an open problem under active investigation \cite{Homma:1997wtc,Berggren:1970wto,Berggren:1996kno,Myo:2011tc,Myo:2023heu}.

Finally, the RMS radius ($r_{\sm{RMS}}$) is defined via the c-product expectation value \cite{Shimizu:2016rrd}:
\begin{equation}
\label{eq:rms_radius}
r^2_{\sm{RMS}}=(\psi|r^2|\psi) = \sum_{i} \int r^2 \psi_{i}(\mathbf{r})^2 d^3\mathbf{r},
\end{equation}

\section{Numerical Results and Discussion}\label{results}

\subsection{Channel Selection and Parameter Fixing}

Based on the formalism established above, we proceed to investigate the mass spectrum of the charmonium-like states. In Tab.\ref{tab:thresholds}, we list the low-lying open-flavor meson-meson thresholds considered in this work, using masses from the Particle Data Group (PDG) review~\cite{Workman:2022ynf}.

Since the strong interaction conserves parity ($P$) and charge conjugation ($C$), the coupling between the $c\bar{c}$ core and the meson-meson channels is allowed only when they share the same $J^{PC}$ quantum numbers. Each meson-meson configuration with a distinct value of the relative orbital angular momentum $L_{M\bar{M}}$ is treated as a separate channel. Furthermore, we restrict our study to the isoscalar sector ($I=0$), implying $G=C$. The selection rules for the allowed orbital angular momenta are summarized in Table~\ref{tab:quantum_numbers}.

\begin{table}[h]
\centering
\caption{Open-charm meson-meson channels and their threshold masses $T_n$ (MeV) used in the calculations~\cite{Workman:2022ynf}.}
\label{tab:thresholds}
\begin{tabular}{cc}
\hline\hline
Channel & Threshold $T_n$ (MeV) \\
\hline
$D\overline{D}$ & 3730 \\
$D\bar{D}^*(2007)$ & 3872 \\
$D_s^+D_s^-$ & 3937 \\
$D^*(2007)\bar{D}^*(2007)$ & 4014 \\
$D_s^+\bar{D}_s^{*-}$ & 4080 \\
$D_s^{*+}\bar{D}_s^{*-}$ & 4224 \\
\hline\hline
\end{tabular}
\end{table}

To perform the numerical calculation, the model parameters need to be determined. Our potential model involves a total of seven parameters. The first four parameters ($m_c, \alpha_s, b, \sigma$) characterize the $c\bar{c}$ interaction. A technical issue arises from the $1/r^3$ singularity present in the spin-orbit and tensor potentials as $r \to 0$. Instead of introducing a hard cutoff radius, we regularize the singular short-distance terms using a Gaussian smearing function. The divergent $1/r^3$ term is replaced by a smooth form \cite{Wu:2024ocq}:
\begin{equation}
\frac{1}{r^3} \to \frac{(1 - e^{-\sigma_1^2 r^2})^2}{r^3},
\label{eq:regularization}
\end{equation}
where $\sigma_1$ is the regularization parameter introduced to smear out the short-range singularity.

\begin{table}[h]
\centering
\caption{Allowed orbital angular momenta ($l$) for the $c\bar{c}$ core and various meson-meson decay channels corresponding to specific $J^{PC}$ quantum numbers.}
\label{tab:quantum_numbers}
\begin{tabular}{ccccc}
\hline\hline
$J^{PC}$ & $l_{c\bar{c}}$ & $l_{D_{(s)}\bar{D}_{(s)}}$ & $l_{D_{(s)}\bar{D}_{(s)}^*}$ & $l_{D_{(s)}^*\bar{D}_{(s)}^*}$ \\
\hline
$1^{--}$ & 0, 2 & 1 & 1 & 1, 3 \\
$2^{++}$ & 1, 3 & 2 & 2 & 0, 2, 4 \\
$1^{++}$ & 1 & - & 0, 2 & 2 \\
$0^{++}$ & 1 & 0 & - & 0, 2 \\
$0^{-+}$ & 0 & - & 1 & 1 \\
$1^{+-}$ & 1 & - & 0, 2 & 0, 2 \\
\hline\hline
\end{tabular}
\end{table}

For the potential parameters ($m_c, \alpha_s, b, \sigma$), we adopt the values from Ref.~\cite{PhysRevD.95.034026}, which were determined by fitting the spectrum of 12 well-established charmonium states. The values are listed in Table~\ref{tab:parameters}.

The remaining two parameters, $\rho$ and $\Delta$, characterize the mixing range and strength, respectively. The range parameter $\rho$ is fixed at $0.3$~fm, consistent with earlier diabatic studies~\cite{PhysRevD.102.074002}. The mixing strength $\Delta$ is treated as a free parameter and determined by reproducing the mass of the exotic candidate $\chi_{c1}(3872)$, yielding $\Delta = 0.110$~GeV. We adopt a single, universal value of $\Delta$ for all partial waves and spin channels. It should be noted that a complete B-O treatment would introduce channel-dependent angular-momentum coefficients in the quarkonium--meson-pair couplings~\cite{Bruschini:2024fyj}. However, since our primary goal is to extend the previous diabatic framework~\cite{Zhang:2025bex} by incorporating the complex scaling method, keeping the same universal mixing prescription allows a direct comparison with the earlier bound-state calculation, thereby isolating the effects of the continuum treatment. Consequently, the pole positions and component weights reported here are
obtained within this universal-coupling approximation.

We also note that the lattice QCD analysis of Ref.~\cite{Bulava:2024jpj}
indicates that the mixing potential is compatible with a constant over the
$1.0$--$1.5$~fm region. To assess the sensitivity of our results to this
radial dependence, we have repeated the calculation with a broader profile,
$\rho = 0.7$~fm, which yields an essentially constant mixing potential over
this region ($\lesssim 6\%$ variation). After recalibrating $\Delta$ to the
$\chi_{c1}(3872)$ mass for this new profile, we find that the absolute mass
shifts are within $10$~MeV and the width variations are generally moderate.
This indicates that, once $\Delta$ is recalibrated for each profile, our main
conclusions are insensitive to the specific choice of the mixing-potential
width.

\begin{table}[h]
\centering
\caption{Model parameters used in the calculation.}
\label{tab:parameters}
\begin{tabular}{cc|cc}
\hline\hline
Parameter & Value & Parameter & Value \\
\hline
$m_c$ & 1.4830 GeV & $\sigma_1$ & 1.5 GeV \\
$\alpha_s$ & 0.5461 & $\rho$ & 0.3 fm \\
$b$ & 0.1425 GeV$^2$ & $\Delta$ & 0.110 GeV \\
$\sigma$ & 1.1384 GeV & & \\
\hline\hline
\end{tabular}
\end{table}

Finally, the coupled-channel Schrödinger equation is solved numerically using the Gaussian Expansion Method (GEM)~\cite{Hiyama:2003cu,Hiyama:2018ivm}. In this variational approach, the radial wave functions are expanded in terms of Gaussian basis functions with range parameters distributed in a geometric progression, which ensures an accurate description from short to long distances.

\subsection{Results and Discussion}
\begin{figure*}[htbp]
    \centering
    \includegraphics[width=0.95\textwidth]{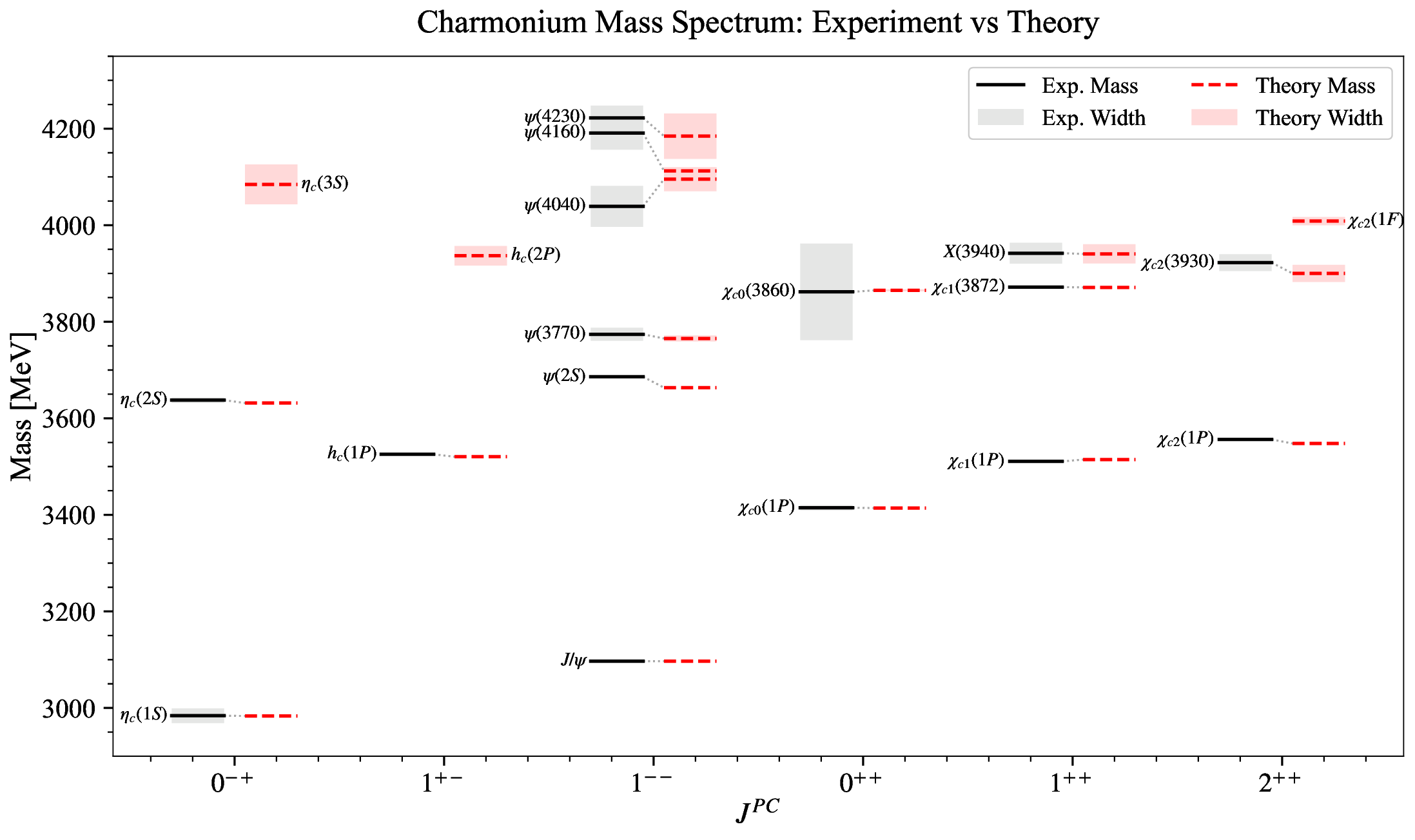}
    \caption{\label{fig:spectrum_comparison}
    Comparison of the theoretical charmonium mass spectrum (red dashed lines) with experimental data (black solid lines). The experimental values are taken from the PDG~\cite{Workman:2022ynf}. The shaded regions represent the decay widths: the pink bands denote the theoretical widths obtained via the CSM, while the gray bands indicate the experimental widths. States are grouped by their $J^{PC}$ quantum numbers along the horizontal axis.}
\end{figure*}
By solving the coupled-channel Schrödinger equation within the framework of the CSM, we obtain the complex energy eigenvalues of the charmonium-like states. The real part of the eigenvalue corresponds to the mass, while the imaginary part relates to the decay width. The numerical results—including the calculated masses, decay widths, and structural properties such as the channel component weights and complex root-mean-square radii ($r_{\sm{RMS}}$)—are summarized in Tab.~\ref{tab:results_summary}. 

To provide an overview of our results, the calculated mass spectrum and decay widths for the charmonium-like states are summarized in Fig.~\ref{fig:spectrum_comparison}, where they are compared with the corresponding experimental data from the PDG~\cite{Workman:2022ynf}. In the following, we discuss the results in detail, organized by $J^{PC}$ quantum numbers.

\begin{figure}[htbp]
    \centering
    \includegraphics[width=0.9\columnwidth]{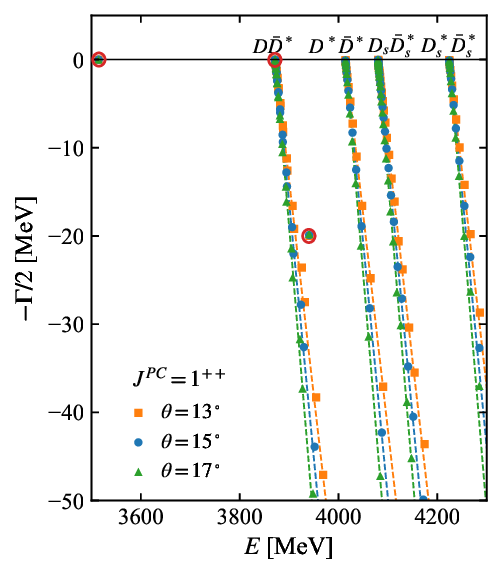}
    \caption{\label{fig:spectrum_1pp} 
    Calculated complex energy eigenvalues for the $J^{PC}=1^{++}$ sector with scaling angles $\theta$ ranging from $13^\circ$ to $17^\circ$.} 
\end{figure}

We begin our discussion with the $J^{PC}=1^{++}$ sector. The calculated complex energy eigenvalues are displayed in Fig.~\ref{fig:spectrum_1pp}. By tracking the stationary points along the energy trajectories as a function of the complex scaling angle $\theta$, we isolate three physical poles. The red circles in the figure mark the $\theta$-stable eigenvalues corresponding to resonance poles. 

The lowest-lying pole is located at a mass of approximately $3515$ MeV. This state is naturally identified as the $\chi_{c1}(1P)$. Since no strong decay channels are kinematically accessible, its width is expected to be zero if weak and electromagnetic decays are neglected. Our numerical result yields a vanishingly small width for this state, fully consistent with its nature as a bound state in our model.

\begin{table*}[htbp]
\centering
\caption{Calculated mass ($M_{th}$), decay width ($\Gamma_{th}$), the $c\bar{c}$ component weight($P_{c\bar{c}}$), and the complex root-mean-square radius ($r_{\sm{RMS}}$) for the charmonium-like states. The dominant component(DC) in the wave function is also listed. The experimental values ($M_{exp}$ and $\Gamma_{exp}$) are taken from the PDG~\cite{Workman:2022ynf}. Masses and widths are given in MeV, while radii are expressed in fm.}
\label{tab:results_summary}
\renewcommand{\arraystretch}{1.3}
\setlength{\tabcolsep}{4pt}

\begin{tabular}{cccccccccc}
\hline\hline
$J^{PC}$ & $M_{th}$ & $\Gamma_{th}$ & $Re P_{c\bar{c}}$ (\%) & $Im P_{c\bar{c}}$ (\%) & $r_{\sm{RMS}}$ (fm) & DC & Assignment & $M_{exp}$ & $\Gamma_{exp}$ \\
\hline
$0^{-+}$ & $2983$ & $0.0$ & $100$ & $0$ & $0.34 - 0.12i$ & $c\bar{c}$ & $\eta_c(1S)$ & $2983.9 \pm 0.4$ & $32.0 \pm 0.7$ \\
         & $3632$ & $0.1$ & $99$ & $0$ & $0.79 - 0.29i$ & $c\bar{c}$ & $\eta_c(2S)$ & $3637.5 \pm 1.1$ & $11.3^{+3.2}_{-2.9}$ \\
         & $4085$ & $82.6$ & $7$ & $29$ & $3.80 - 0.23i$ & $D^*\bar{D}^*$ & -- & -- & -- \\
\hline
$1^{--}$ & $3097$ & $0.0$ & $100$ & $0$ & $0.40 - 0.12i$ & $c\bar{c}$ & $J/\psi(1S)$ & $3096.9 \pm 0.006$ & $0.093 \pm 0.003$ \\
         & $3663$ & $0.0$ & $93$ & $0$ & $0.87 - 0.27i$ & $c\bar{c}$ & $\psi(2S)$ & $3686.10 \pm 0.01$ & $0.294 \pm 0.008$ \\
         & $3765$ & $12.3$ & $96$ & $20$ & $0.64 + 0.15i$ & $c\bar{c}$ & $\psi(3770)$ & $3773.7 \pm 0.4$ & $27.2 \pm 1.0$ \\
         & $4095$ & $49.6$ & $34$ & $7$ & $1.49 + 1.11i$ & $D^*\bar{D}^*$ & $\psi(4040)$ & $4039 \pm 1$ & $80 \pm 10$ \\
         & $4112$ & $14.7$ & $89$ & $6$ & $1.33 - 0.26i$ & $c\bar{c}$ & $\psi(4160)$ & $4191 \pm 5$ & $70 \pm 10$ \\
         & $4184$ & $94.1$ & $25$ & $27$ & $2.00 - 1.63i$ & $D_s\bar{D}_s^*$ & $\psi(4230)$ & $4222.7 \pm 2.6$ & $49 \pm 8$ \\
\hline
$0^{++}$ & $3414$ & $0.0$ & $100$ & $0$ & $0.56 - 0.15i$ & $c\bar{c}$ & $\chi_{c0}(1P)$ & $3414.7 \pm 0.3$ & $10.8 \pm 0.6$ \\
         & $3865$ & $3.2$ & $77$ & $8$ & $0.99 - 0.24i$ & $c\bar{c}$ & $\chi_{c0}(3860)$ & $3862^{+26}_{-32}$ & $201^{+154}_{-67}$ \\
\hline
$1^{++}$ & $3515$ & $0.0$ & $100$ & $0$ & $0.65 - 0.17i$ & $c\bar{c}$ & $\chi_{c1}(1P)$ & $3510.67 \pm 0.05$ & $0.84 \pm 0.04$ \\
         & $3871$ & $0.03$ & $11$ & $0$ & $3.85 - 1.07i$ & $D\bar{D}^*$ & $\chi_{c1}(3872)$ & $3871.65 \pm 0.06$ & $1.19 \pm 0.21$ \\
         & $3941$ & $39.9$ & $56$ & $30.3$ & $1.46 - 1.24i$ & $c\bar{c}$ & $X(3940)$ & $3942 \pm 9$ & $43^{+28}_{-18}$ \\
\hline
$1^{+-}$ & $3520$ & $0.0$ & $100$ & $0$& $0.65 - 0.20i$ & $c\bar{c}$ & $h_c(1P)$ & $3525.38 \pm 0.11$ & $0.7 \pm 0.4$ \\
         & $3937$ & $40.5$ & $56$ & $30$& $0.71 - 1.12i$ & $c\bar{c}$ & -- & -- & -- \\
\hline
$2^{++}$ & $3548$ & $0.0$ & $99$ & $0$ & $0.68 - 0.25i$ & $c\bar{c}$ & $\chi_{c2}(1P)$ & $3556.17 \pm 0.07$ & $1.97 \pm 0.09$ \\
         & $3900$ & $35.4$ & $48$ & $-10$ & $4.65 - 0.42i$ & $D\bar{D}^*$ & $\chi_{c2}(3930)$ & $3922.5 \pm 1.0$ & $35.2 \pm 2.2$ \\
         & $4009$ & $16.9$ & $107$ & $-7$ & $0.89 - 0.08i$ & $c\bar{c}$ & -- & -- & -- \\
\hline\hline
\end{tabular}
\end{table*}

\begin{figure}[htbp]
    \centering
    \includegraphics[width=0.9\columnwidth]{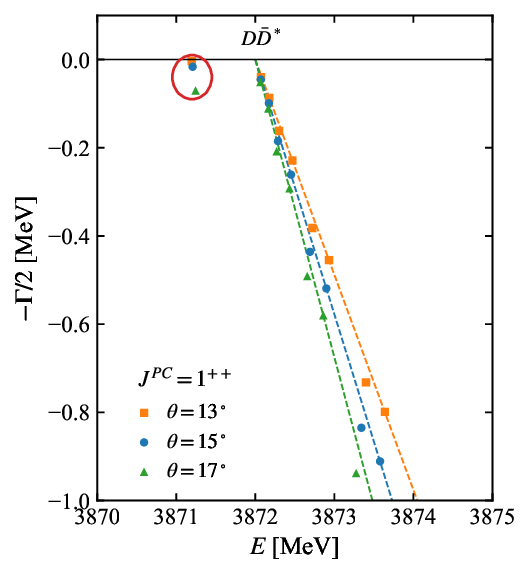}
    \caption{\label{fig:pole_3872} 
    Zoomed-in view of the complex energy plane for the $J^{PC}=1^{++}$ sector near the $D\bar{D}^*$ threshold. The precise pole position of the $\chi_{c1}(3872)$ is marked, demonstrating its very close proximity to the real axis and the $D\bar{D}^*$ threshold.}
\end{figure}
\begin{figure}[htbp]
    \centering
    \includegraphics[width=0.9\columnwidth]{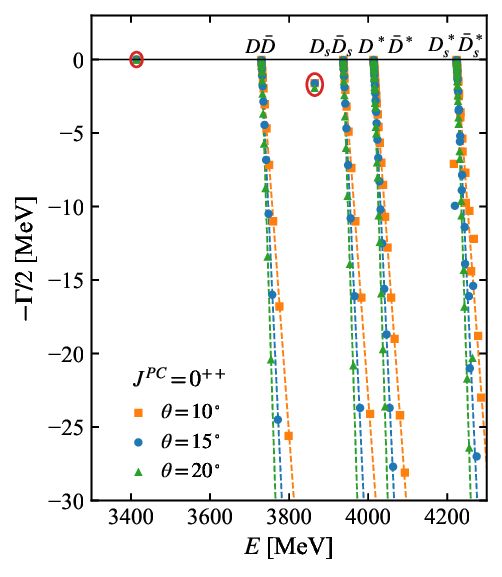}
    \caption{\label{fig:spectrum_0pp} 
    Calculated complex energy eigenvalues for the $J^{PC}=0^{++}$ sector with scaling angles $\theta$ ranging from $10^\circ$ to $20^\circ$.}
\end{figure}

The second pole observed in the $J^{PC}=1^{++}$ sector is of particular interest. It is located in the immediate vicinity of the $D\bar{D}^*$ threshold. Given its proximity to the threshold, Fig.~\ref{fig:pole_3872} provides a magnified view of the complex energy plane, clearly illustrating the trajectory of the pole as the scaling angle $\theta$ varies. As shown in the figure, the pole exhibits excellent stability as $\theta$ varies, confirming its nature as a physical resonance. 

The extracted pole mass agrees well with that of the experimentally observed $\chi_{c1}(3872)$ (also known as $X(3872)$). The mass of this pole is used in the fit to constrain the parameters of the mixing potential. The imaginary part of the eigenenergy is remarkably small, corresponding to a very narrow decay width, which is in good agreement with the experimental upper limit. Furthermore, the component analysis shows that the real part of the $D\bar D^*$ component weight reaches $88.4\%$, while its imaginary part is only $-0.1\%$, indicating that the state is dominated by the $D\bar D^*$ channel.  The corresponding complex RMS radius is $3.85-1.07i$ fm; although it should not be interpreted as an ordinary real-valued radius, its sizable real part suggests an extended spatial structure. Based on this large $D\bar{D}^*$ weight and extended spatial size, we interpret this state as a loosely bound hadronic molecule generated dynamically by the channel coupling, rather than a conventional compact charmonium state. This result is consistent with several theoretical studies~\cite{Tornqvist:2004qy,Guo:2013sya,Braaten:2003he,Swanson:2003tb,
PhysRevD.72.034010,Gamermann:2009fv,Ortega:2009hj,
RevModPhys.90.015004,Brambilla:2019esw,Brambilla:2024zvx}.

 We also identify a pole with a mass of $3940$ MeV and a width of $40$ MeV in the $1^{++}$ sector. Although no charmonium-like state with $J^{PC}=1^{++}$ has been reported in this energy region to date, the experimentally observed $X(3940)$ state~\cite{Belle:2005lik,LHCb:2024vfz}, whose quantum numbers remain undetermined, stands out as a likely candidate. According to the PDG data~\cite{Workman:2022ynf}, the mass and width of the $X(3940)$ are $3942\pm 9$ MeV and $43^{+28}_{-18}$ MeV, respectively. Our calculated results ($M \approx 3940$ MeV, $\Gamma \approx 40$ MeV) are in excellent agreement with these experimental values.

Furthermore, this assignment is supported by the decay channels observed in experiments. The $X(3940)$ has been observed to decay primarily into $D\bar{D}^*$, while the $D\bar{D}$ decay mode has not been reported. Under our proposed $J^{PC}=1^{++}$ assignment, the coupling of this state to the $D\bar{D}$ channel is forbidden by parity and angular momentum conservation, whereas the coupling to $D\bar{D}^*$ is allowed and dominant. This consistency between the calculated properties and the experimental decay characteristics supports identifying the $X(3940)$ as a $J^{PC}=1^{++}$ state, which aligns with the results reported in the literature \cite{Ortega:2009hj,Zhou:2017dwj}. 
Further component analysis reveals that the real parts of the $c\bar{c}$ and $D\bar{D}^*$ component weights are $55.8\%$ and $40.9\%$, respectively, while the corresponding imaginary parts are $30.3\%$ and $-31.2\%$. The sizable imaginary parts indicate that these complex quantities cannot be interpreted as ordinary probabilities. Nevertheless, the comparable real parts of the component weights from the $c\bar{c}$ and $D\bar{D}^*$ channels suggest strong mixing between the charmonium and meson-meson configurations. Consequently, this state is best interpreted as a mixture of charmonium and hadronic molecular components.

\begin{figure}[htbp]
    \centering
    \includegraphics[width=0.9\columnwidth]{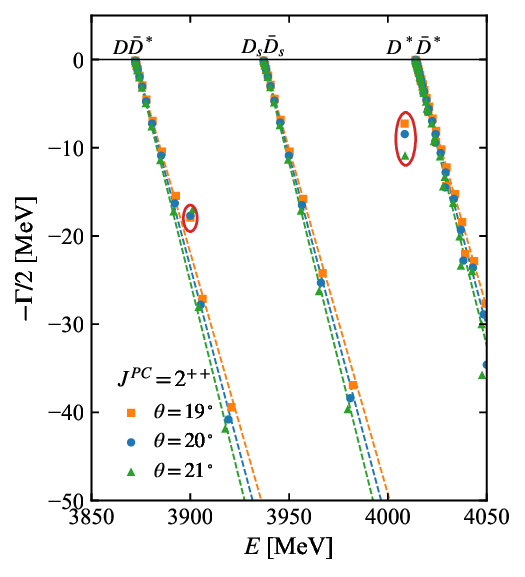}
    \caption{\label{fig:spectrum_2pp} 
   Calculated complex energy eigenvalues in the $J^{PC}=2^{++}$ sector as a function of the scaling angle $\theta$ ($19^\circ \leq \theta \leq 21^\circ$).}
\end{figure}
Turning to the $J^{PC}=0^{++}$ sector, we identify two poles as shown in Fig. \ref{fig:spectrum_0pp}. The lower-lying pole is located at $M \approx 3415$ MeV, corresponding to the $\chi_{c0}(1P)$ state, which is consistent with predictions from quenched potential models and experimental measurements\cite{Eichten:1974af,PhysRevD.17.3090,Godfrey:1985xj,barnes2005higher,eichten2008quarkonia,ParticleDataGroup:2024cfk}. The second pole is found at $M = 3865$ MeV with a narrow width of $\Gamma = 4$ MeV. Our analysis of channel contributions reveals that the $c\bar{c}$ component constitutes approximately $77\%$ of its wave function.

The identification of the first excited scalar charmonium state, $\chi_{c0}(2P)$, remains a subject of debate, with two primary candidates: the $\chi_{c0}(3860)$ and the $\chi_{c0}(3915)$. According to the PDG averages~\cite{Workman:2022ynf}, the mass and width of the $\chi_{c0}(3860)$ are $3862^{+26+40}_{-32-13}$ MeV and $201^{+154+88}_{-67-82}$ MeV, respectively. Our calculated mass of $3865$ MeV is in excellent agreement with the experimental mass of the $\chi_{c0}(3860)$, supporting its assignment as the $\chi_{c0}(2P)$ state. However, a discrepancy arises regarding the decay width: the experimental data suggest a broad state ($\Gamma \sim 200$ MeV), whereas our calculation yields a narrow width of approximately $4$ MeV. Notably, our result is consistent with several other theoretical studies~\cite{Man:2024mvl,Zhou:2017dwj,barnes2005higher}, which also predicted a relatively narrow width for the $\chi_{c0}(2P)$ state based on various theoretical frameworks. This tension between the broad experimental width and the narrower theoretical predictions indicates that the nature of the experimentally observed $\chi_{c0}(3860)$ may be more complex than currently understood. Therefore, further theoretical and experimental investigations are required to fully clarify the nature of the $\chi_{c0}(2P)$ resonance and accurately determine its total width.

The other candidate, the $\chi_{c0}(3915)$, was primarily observed in the $\gamma\gamma \to \omega J/\psi$ process~\cite{Belle:2004lle, LHCb:2022aki}. According to Refs.~\cite{ortega2018charmonium,PhysRevLett.115.022001,Guo:2012tv,Duan:2020tsx}, identifying this state as the $\chi_{c0}(2P)$ poses a significant theoretical challenge. As a $P$-wave charmonium state located above the open-charm threshold, the $\chi_{c0}(2P)$ is expected to decay predominantly via the OZI-allowed $D\bar{D}$ channel, resulting in a large total width. However, experimental observations present a contradictory picture: the $X(3915)$ exhibits a significant branching fraction into the hidden-charm channel $\omega J/\psi$, while no significant signal has been observed for the open-charm $D\bar{D}$ decay. Consequently, the interpretation of the $X(3915)$ as the conventional $\chi_{c0}(2P)$ remains controversial.

In the present coupled-channel calculation, we have included only open-flavor meson-meson thresholds, neglecting hidden-charm channels such as $\omega J/\psi$. Notably, our results do not yield a pole corresponding to the $X(3915)$. This absence is consistent with the anomalous nature of the $X(3915)$, suggesting that this state cannot be described as a conventional $c\bar{c}$ state coupled mainly to open-charm channels. Its formation likely requires the explicit inclusion of hidden-charm dynamics or other mechanisms.

\begin{figure}[H]
    \centering
    \includegraphics[width=0.9\columnwidth]{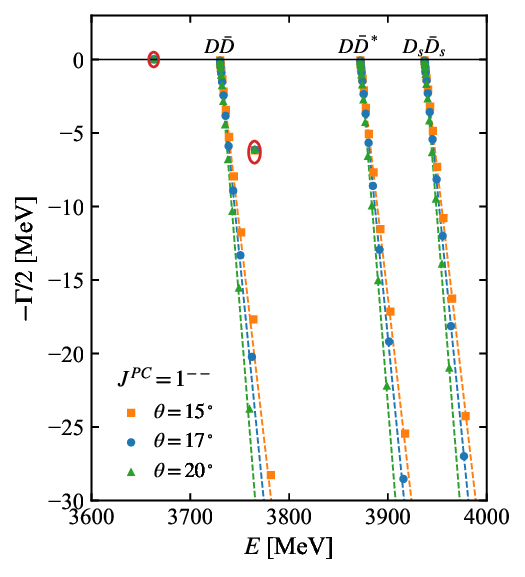}
    \caption{\label{fig:spectrum_1mm1} 
    Calculated complex energy eigenvalues for the $J^{PC}=1^{--}$ sector in the energy region from $3600$ to $4000$ MeV, with scaling angles $\theta$ ranging from $15^\circ$ to $20^\circ$.}
\end{figure}
We next turn our attention to the $J^{PC}=2^{++}$ sector. The excited states and their pole structures in the complex energy plane are illustrated in Fig.~\ref{fig:spectrum_2pp}.

The first pole in this sector is located at $M \approx 3900$ MeV with $\Gamma \approx 36$ MeV. This result is consistent with the experimentally established $\chi_{c2}(3930)$ (also denoted $\chi_{c2}(2P)$), which has a mass of $3922.5 \pm 1.0$ MeV, a width of $35.2 \pm 2.2$ MeV~\cite{Workman:2022ynf}, and predominantly decays into $D\bar{D}$. Component analysis reveals that the real part of the $c\bar{c}$ weight is approximately $48\%$, indicating that the meson-meson components make a substantial contribution to the internal structure of this state.
This significant configuration mixing underscores the importance of coupled-channel dynamics, which are indispensable for accurately reproducing the mass and width of the $\chi_{c2}(3930)$.

In addition to the $\chi_{c2}(3930)$, our calculation predicts a second pole at $M = 4009$ MeV with a narrow width of $\Gamma = 16$ MeV. To date, no $J^{PC}=2^{++}$ charmonium-like state has been observed in this mass region. The analysis reveals that the real part of the $c\bar{c}$ component weight for this state reaches approximately $107.3\%$. Although a value exceeding unity might seem surprising, it is a natural consequence of the non-Hermitian nature of complex resonance poles: the component weights are generally complex-valued and not constrained to the interval [0, 1], thus lacking a rigorous probabilistic interpretation\cite{Lin:2023ihj,Song:2025xnx,Wu:2024qve,Myo:2021xqp}. Nevertheless, its proximity to unity primarily reflects the dominant $c\bar{c}$ character of this state. Given its substantial $c\bar{c}$ core and predominantly $F$-wave configuration, we tentatively assign this resonance to the $\chi_{c2}(1F)$ state. An experimental search for a tensor state near 4.0 GeV would provide a crucial test of our coupled-channel framework.

\begin{figure}[htbp]
    \centering
    \includegraphics[width=0.9\columnwidth]{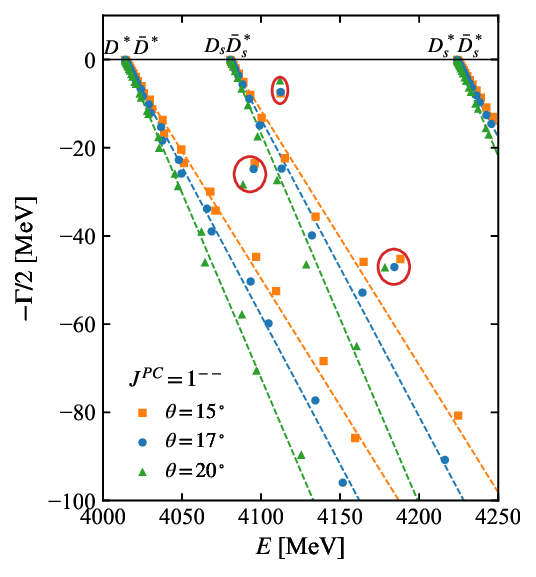}
    \caption{\label{fig:spectrum_1mm2} 
    Calculated complex energy eigenvalues for the $J^{PC}=1^{--}$ sector between $4000$ and $4250$ MeV, with scaling angles $\theta$ ranging from $15^\circ$ to $20^\circ$.}
\end{figure}
\begin{figure}[htbp]
    \centering
    \includegraphics[width=0.9\columnwidth]{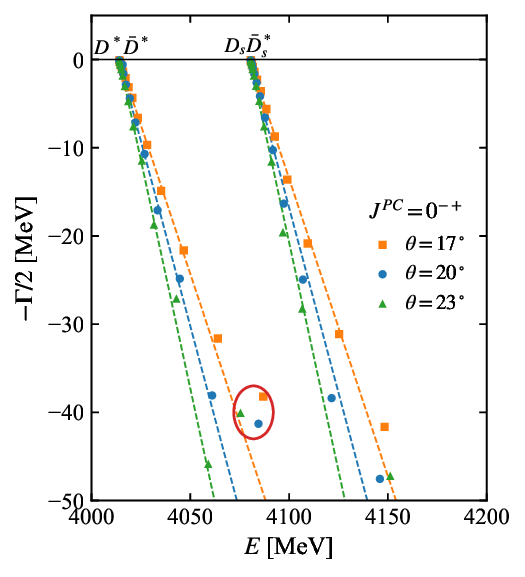}
    \caption{\label{fig:spectrum_0mp1} 
    Calculated complex energy eigenvalues for the $J^{PC}=0^{-+}$ sector in the energy region from $4000$ to $4200$ MeV, with scaling angles $\theta$ ranging from $17^\circ$ to $23^\circ$.}
\end{figure}

\begin{figure}[htbp]
    \centering
    \includegraphics[width=0.9\columnwidth]{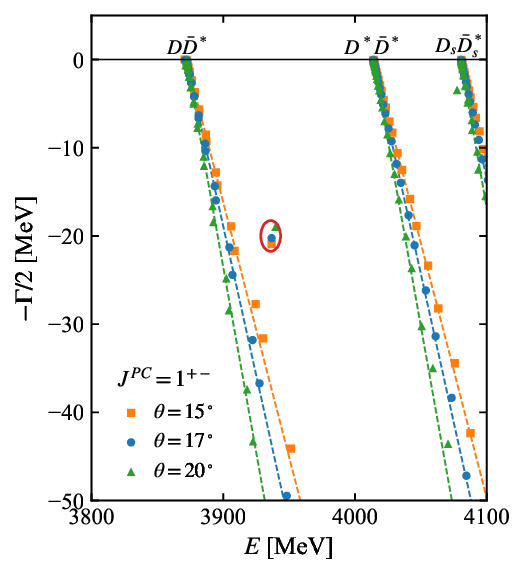}
    \caption{\label{fig:spectrum_1pm1} 
    Calculated complex energy eigenvalues for the $J^{PC}=1^{+-}$ sector in the energy range $3800-4100$ MeV, with scaling angles $\theta$ ranging from $15^\circ$ to $20^\circ$.}
\end{figure}
Turning to the vector meson sector ($J^{PC}=1^{--}$), we find that the first excited pole above the $J/\psi(1S)$ ground state is located at $M \approx 3663$ MeV, as shown in Fig.~\ref{fig:spectrum_1mm1}. Positioned well below the lowest open-charm threshold, this pole corresponds to a bound state with vanishing width within our coupled-channel framework. We identify this state as the $\psi(2S)$, which has an experimental mass of $3686.097 \pm 0.010$ MeV~\cite{Workman:2022ynf}. 

The second pole is located at $M = 3765$ MeV with a width of $\Gamma = 12.4$ MeV. Component analysis further indicates that this state is dominated by a $D$-wave $c\bar{c}$ configuration, allowing us to identify it as the well-established $\psi(3770)$ 
resonance. While the calculated mass is in good agreement with the experimental value 
($3773.7 \pm 0.4$~MeV)~\cite{Workman:2022ynf}, the predicted width is noticeably 
smaller than the world average ($\Gamma \approx 27.2$~MeV).
 
The complex poles in the higher-energy region of the $1^{--}$ sector are presented in Fig.~\ref{fig:spectrum_1mm2}. The first pole in this region is located at $M = 4095$~MeV with a width of $\Gamma = 49.6$~MeV. Although the calculated mass is somewhat higher and the width narrower than their corresponding experimental values, the deviations remain moderate. Component analysis reveals that the $c\bar{c}$ weight is only $34\%$, indicating that this state possesses a substantial $D^*\bar{D}^*$ molecular component. Given its mass and strong coupling to the continuum, this state can be regarded as the theoretical counterpart to the experimentally observed $\psi(4040)$.

The second pole in this region is located at $M = 4112$~MeV with a width of $\Gamma = 14$~MeV. This state can be identified with the $\psi(4160)$ resonance; however, its calculated width is notably narrower than the experimental value $\Gamma = 70 \pm 10$ MeV. The $c\bar{c}$ component dominates this state ($89\%$). Furthermore, a detailed analysis reveals that within the $c\bar c$ component, the $S$-wave and $D$-wave contributions are comparable, demonstrating that it is a conventional $c\bar{c}$ charmonium state with significant $S$-$D$ mixing.

The third pole in this region is located at $M = 4184$~MeV with a broad width of $\Gamma = 90$~MeV. This resonance is tentatively identified with the $\psi(4230)$ state, whose experimentally measured mass and width are $M_{\rm exp}=4222.7\pm2.6$~MeV and $\Gamma_{\rm exp}=49\pm8$~MeV, respectively. Component analysis reveals that the $D_s\bar{D}_s^*$ component weight is complex, with a sizable real part of $74.0\%$ and a substantial imaginary part of $-61.9\%$. As discussed above, due to the non-Hermitian nature of complex resonance poles, such weights lack a strict probabilistic interpretation. Nevertheless, the large real part strongly suggests a significant $D_s\bar{D}_s^*$ molecular component.

In the vector sector ($1^{--}$), the number of states predicted by our model precisely matches the number of experimentally observed vector charmonium states in this energy region. Overall, for this specific quantum number, our framework successfully reproduces the observed state multiplicity, despite minor quantitative discrepancies in the calculated masses and widths. These deviations may arise from the specific parametrization of the model or the neglect of additional decay channels.

Finally, we present our results for the $J^{PC} = 0^{-+}$ and $1^{+-}$ sectors.

In the $0^{-+}$ sector, below the lowest open-flavor threshold, we obtain two bound states. The ground state has a mass of $2983$ MeV, while the first excited state is at $3632$ MeV. These values are in excellent agreement with the well-established experimental masses of the $\eta_c(1S)$ and $\eta_c(2S)$ states, respectively. At higher energies, as illustrated in Fig.~\ref{fig:spectrum_0mp1}, we identify a resonant pole at $M \approx 4085$ MeV with a width of $\Gamma \approx 82$ MeV.

In the $1^{+-}$ sector, our calculation yields a ground state mass of $3520$ MeV, which can be identified with the $h_c(1P)$ meson. Furthermore, we locate an excited pole at $M \approx 3937$ MeV with a width of $\Gamma \approx 40$ MeV, as shown in Fig.~\ref{fig:spectrum_1pm1}.

Given that experimental information regarding these higher-mass states is currently limited, the poles found at $4085$ MeV ($0^{-+}$) and $3937$ MeV ($1^{+-}$) serve as theoretical predictions. Future experimental measurements will be crucial to verify the existence of these resonances.

\section{Summary}

In this work, we present a comprehensive study of charmonium-like states within a diabatic approach. By explicitly including open-flavor meson-meson channels, we provide a unified framework for describing both conventional charmonia and exotic candidates. Furthermore, the application of the CSM enables us to solve bound-state and resonant-state problems on an equal footing, without the need to introduce additional parameters.

For the low-lying charmonium states below the open-flavor threshold, including the $J/\psi(1S)$, $\eta_c(1S)$, $h_c(1P)$, and $\chi_{cJ}(1P)$, our calculated masses and widths are in excellent agreement with experimental data. These results confirm the reliability of our potential parameters and the validity of the coupled-channel framework in describing the $c\bar{c}$ system.

For the exotic candidates, our framework naturally accommodates the $X(3872)$ as a hadronic molecular state. Furthermore, the theoretical predictions for the mass, width, and decay properties of the $X(3940)$ are consistent with a $J^{PC}=1^{++}$ assignment.

For the controversial $\chi_{c0}(2P)$ candidate, our calculated mass supports its identification as the $\chi_{c0}(3860)$. Although the theoretical width is narrower than the current experimental central value, it remains consistent with other theoretical predictions, suggesting that the experimental width may require further investigation.

In the $2^{++}$ sector, our calculated mass and width exhibit remarkable agreement with experimental measurements of the $\chi_{c2}(3930)$. Furthermore, component analysis demonstrates that the $\chi_{c2}(3930)$ is not a pure charmonium state, but rather a state strongly dressed by meson-meson clouds.

In the vector sector ($1^{--}$), our model successfully reproduces the number of highly excited states observed in experiments, although minor quantitative discrepancies in masses and widths remain.

However, we note that the current calculation is restricted to open-charm meson-meson channels. Experimental observations indicate that several exotic candidates, such as the $X(3915)$, $\chi_{c1}(4140)$, and $\chi_{c1}(4274)$, exhibit significant or dominant decay rates into hidden-charm modes (e.g., $J/\psi \phi$ and $J/\psi \omega$). The absence of these states in our present spectrum suggests that their formation and decay dynamics are likely governed by mechanisms involving hidden-charm channels, which are not yet included in our model. Therefore, a natural and necessary extension of this work is to incorporate hidden-charm channels (such as $\omega J/\psi$ and $\phi J/\psi$) into the coupled-channel formalism. Such an extension will be crucial for a comprehensive understanding of these exotic states and for completing the physical picture of the charmonium spectrum in the higher-mass region.

\begin{acknowledgements}
We acknowledge support from the Natural Science Basic Research Program of Shaanxi Province under Grant Nos. 2024JC-YBMS-010 and 2026JC-YXQN-004.

\end{acknowledgements}

%

\end{document}